# Superhabitable Planets Around Mid-Type K Dwarf Stars Enhance Simulated JWST Observability and Surface Habitability

**Iva Vilović[1]\*, Jayesh Goyal[2], René Heller[3], and Fanny Marie von Schauenburg[1]**

[1]Astrobiology Research Group, Zentrum für Astronomie und Astrophysik, Technische Universität Berlin, 10623 Berlin, Germany.
[2]School of Earth & Planetary Sciences (SEPS), National Institute of Science Education & Research (NISER), Bhubaneshwar, Odisha - 752050, India.
[3]Max-Planck-Institut für Sonnensystemforschung, 37077 Göttingen, Germany.
**\*Corresponding author. Email: iva.vilovic@campus.tu-berlin.de**

**In our search for life beyond the Solar System, certain planetary bodies may be more conducive to life than Earth. However, the observability of these 'superhabitable' planets in the habitable zones around K dwarf stars has not been fully modeled. This study addresses this gap by modeling the atmospheres of superhabitable exoplanets. We employed the 1D model *Atmos* to define the superhabitable parameter space, *POSEIDON* to calculate synthetic transmission spectra, and *PandExo* to simulate JWST observations. Our results indicate that planets orbiting mid-type K dwarfs, receiving 80% of Earth's solar flux, are optimal for life. These planets sustain temperate surfaces with moderate $CO_2$ levels, unlike those receiving 60% flux, where necessarily higher $CO_2$ levels could hinder biosphere development. Moreover, they are easier to observe, requiring significantly fewer transits for biosignature detection compared to Earth-like planets around Sun-like stars. For instance, detecting biosignature pairs like oxygen and methane from 30 parsecs would require 150 transits (43 years) for a superhabitable planet, versus over 1700 transits (~1700 years) for Earth-like planets. While such observation times lie outside of JWST mission timescales, our study underscores the necessity of next-generation telescopes and provides valuable targets for future observations with, for example, the ELT.**

Keywords: astrobiology -- methods: numerical -- stars: K dwarfs

**Abbreviations: DDT**, Director's Discretionary Time; **DIVFrms**, divergence from root mean square; **ELT**, Extremely Large Telescope; **ESI**, Earth Similarity Index; **HITEMP**, high-temperature molecular spectroscopic database; **HITRAN**, high-resolution transmission molecular absorption database; **HWO**, Habitable Worlds Observatory; **HZ**, habitable zone; **JWST**, James Webb Space Telescope; **KOBE**, Kepler Observatory of Biosignature Experiments; **LBOUND**, lower boundary condition; **MBOUND**, upper boundary condition; **ME**, modern Earth; **MIRI LRS**, Mid-Infrared Instrument Low-Resolution Spectrometer; **MUSCLES**, Measurements of the Ultraviolet Spectral Characteristics of Low-mass Exoplanetary Systems; **NIRSpec**, Near Infrared Spectrograph; $N_{trans}$, number of transits; **PAL**, present atmospheric level; **PHOENIX**, a spectral library for stellar atmospheres; **POSEIDON**, exoplanet atmosphere model for transmission spectra; **ppm**, parts per million; **PSF**, point spread function; **RH**, relative humidity; **SEPHI**, Statistical-likelihood Exo-Planetary Habitability Index; **S/N**, signal-to-noise ratio; $S_0$, solar constant; **SGFLUX**, constant upward flux; **SH**, superhabitable; **T–P**, temperature-pressure; **Vdep**, deposition velocity; **vmr**, volume mixing ratio; **Z**, stellar metallicity

## 1. Introduction

In the search for life beyond Earth, some exoplanets may be more habitable than our own (Heller and Armstrong 2014). Current studies mainly focus on M dwarf stars–the cooler, smaller, and dimmer counterparts to Sun-like stars (Luger and Barnes 2015, Gillon *et al.* 2017, Wunderlich *et al.* 2019, Claudi *et al.* 2020, Gebauer *et al.* 2021, Battistuzzi *et al.* 2023, Lobo and Shields 2024). However, the high stellar activity and lengthy pre-main sequence phases of M dwarfs pose risks to habitability (Tarter *et al.* 2007, Barnes *et al.* 2009, Luger and Barnes 2015).

K dwarf stars, on the other hand, stand out as promising candidates for superhabitable (SH) worlds due to their long lifespans (50 to 100 billion years), relatively low luminosities (0.08 to 0.6 times the solar luminosity) and cooler temperatures (3700 to 5200 Kelvin) compared to Sun-like stars—which result in more frequent transits of planets in the stellar habitable zone (HZ) and a more favorable brightness contrast (Heller and Armstrong 2014, Cuntz and Guinan 2016, Arney 2019, Schulze-Makuch *et al.* 2020). This contrast improves the signal-to-noise ratio for spectroscopic studies by reducing stellar contamination and enhancing the detectability of key molecular features in planetary atmospheres, particularly in the infrared. K dwarfs, with significantly lower activity levels than M dwarfs, create a more stable environment for orbiting planets (Richey-Yowell *et al.* 2019). Their shorter pre-main sequence phases of under 0.1 billion years, compared to up to 1 billion years for M dwarfs (Baraffe *et al.* 2015), also reduce the risk of atmospheric loss during early planetary development. Additionally, K dwarfs emit less intense radiation with a spectral distribution shifted toward longer, infrared wavelengths compared to solar-type (G-type) stars, resulting in lower X-ray output, which further minimizes atmospheric erosion on orbiting planets (Richey-Yowell *et al.* 2019, 2023). K dwarfs are also more abundant than G-type stars, representing about 12% of the total stellar population compared to 8% for G stars (Arney 2019), and their expected life spans of between 17 and 70 billion years are comparable to, longer or much longer than the age of the Universe. This longevity, along with their relatively stable radiation output, provides a more benign environment for planets to develop and sustain atmospheres on geological and even cosmological timescales. Their relatively low luminosity, as low as one-tenth that of a Sun-like star, offers favorable observational conditions for direct imaging of their planets, if present (Cuntz and Guinan 2016). Notably, the HZs around K dwarfs are, in most of the cases, farther away from the star than the planetary tidal-locking distance. This might be an important asset of Earth-like planets around K dwarf stars compared to M dwarf stars since it would promote stable day-and-night cycles (Kasting *et al.* 1993) and prevent extreme temperature variations on the pro- and anti-stellar hemispheres on the planet.

Recently, there has been increasing interest in K dwarf stars as optimal targets for the search for extraterrestrial life. Studies have modeled K dwarf transmission spectra as well as temperature-pressure (T–P) and chemical abundance profiles (Kaltenegger and Lin 2021); others have analyzed the astrophysical and observational benefits of K dwarf stars (Cuntz and Guinan 2016, Arney 2019); while other studies looked at the influence of orbital obliquity and eccentricity on marine biological activity (Barnett and Olson 2022, Jernigan *et al.* 2023). Furthermore, laboratory experiments have confirmed that photosynthetic organisms can successfully grow under simulated K dwarf star radiation (Vilović *et al.* 2024). However, to our knowledge, no studies have explicitly modeled superhabitable atmospheres based on Earth's varying natural history and astrophysical parameters (Heller and Armstrong 2014, Schulze-Makuch *et al.* 2020, Vilović *et al.* 2023) as well as assessed their observability with state-of-the-art telescopes.

Motivated astrophysically, a superhabitable planet would be slightly larger and more massive than Earth to retain a thicker atmosphere–providing the necessary mass and energy to support a more extensive biosphere–and support plate tectonics (Schulze-Makuch *et al.* 2020). Additionally, a strong magnetic field would protect the planet from harmful radiation, preserving its atmosphere and potential life forms (Lammer *et al.* 2009).

While this study builds on the established paradigm of carbon-based life with water as a supporting solvent, alternative life chemistries could exist. Lifeforms using silicon, sulfur, or nitrogen as building blocks, for example, could hypothetically thrive in environments inhospitable to carbon-based organisms (Bains 2004). Solvents like ammonia or methane, which remain liquid at cryogenic temperatures, might support life under such conditions, opening the possibility of habitats on icy moons or gas giants. However, the slower reaction rates at these low temperatures could mean that life processes, including evolution, might proceed more slowly than in water-based systems. Despite their theoretical viability, non-carbon elements do not form complex, stable molecules as readily as carbon, and non-water solvents lack the chemical versatility and reaction rates conducive to life as we know it. This study focuses on carbon and water because of these advantages and the detectability of carbon-based biosignatures, like oxygen and methane, with current observational technology.

Based on Earth's natural history, particularly the Carboniferous Era around 300 million years ago, which saw the highest biomass production and biological diversity, a superhabitable planet would feature elevated oxygen levels (Vilović *et al.* 2023). This would enable more extensive metabolic networks and support larger organisms (Harrison *et al.* 2010). Additionally, it would have a moderate surface temperature approximately 5 degrees Celsius warmer than present-day Earth and increased atmospheric relative humidity (RH). This is beneficial because the universal solvent properties of water facilitate the transport of essential substances for biophysical processes and ecosystem productivity (Reichardt and Timm 2020).

In terms of the atmospheric composition, key organisms and biological sources affecting Earth's biosphere and their atmospheric signatures are considered. A superhabitable atmosphere would have increased levels of methane ($CH_4$) and nitrous oxide ($N_2O$) due to heightened production by methanogenic microbes, as well as denitrifying bacteria and fungi, respectively (Averill and Tiedje 1982, Wen *et al.* 2017). Furthermore, it would have decreased levels of molecular hydrogen ($H_2$) due to higher enzyme consumption (Lane *et al.* 2010, Greening and Boyd 2020). Lastly, as previously mentioned, molecular oxygen ($O_2$) levels could increase from present-day 21% by volume on Earth to 25% to reflect a thriving photosynthetic biosphere (Schirrmeister *et al.* 2015). Although not universally established, recent laboratory experiments have demonstrated that photosynthetic organisms can grow successfully under simulated K dwarf star radiation, suggesting that specific species may indeed achieve enhanced photosynthesis under these conditions (Vilović *et al.* 2024).

Taking the discussed factors into account, K dwarfs can be regarded as the 'Goldilocks stars' in the search for potentially life-supporting planets, and potentially provide the ideal compromise for both habitability and its observability.

## 2. Methodology

In this work we utilized simulations using three tools to model superhabitable exoplanetary atmospheres:

1) we applied the 1D coupled climate-photochemistry model *Atmos* (Kopparapu *et al.* 2013) to explore the range of conditions that define the superhabitable parameter space and produce temperature-pressure as well as atmospheric abundance profiles for the most prevalent biomarkers as a function of stellar input spectra;

2) we employed the *POSEIDON* forward modeling code (MacDonald and Madhusudhan 2017, MacDonald 2023) to calculate synthetic planetary transmission spectra based on the exoplanetary atmospheres produced in *Atmos*; and lastly

3) we utilized the *PandExo* tool (Batalha *et al.* 2017) to simulate distance-dependent observations of transiting exoplanets with the James Webb Space Telescope (JWST) as well as to determine the lowest number of transits necessary to constrain key spectral features associated with life as we know it on Earth based on the transmission models obtained by *POSEIDON*.

Through this integrated approach, we can assess the observability of potential life on exoplanets, constrain the most observable types of planets, and estimate the optimal timeframes for these observations.

## *2.1. Atmos model description*

*Atmos*, a 1D radiative-convective model, was initially developed by (Kasting *et al.* 1993) and subsequently refined to simulate habitable zones around various star types (Kopparapu *et al.* 2013). It provides a framework for analyzing fundamental atmospheric processes. While 3D models are valuable for investigating complex dynamics like global transport, hydrological cycles, and cloud formations, they require numerous boundary conditions—such as continental distribution, orography, obliquity, rotation rate, and oceanic heat transport—which are largely unknown for potentially habitable rocky extrasolar planets. Therefore, 1D models offer a more practical and focused approach for this study, allowing for a thorough examination of the fundamental atmospheric processes and their implications for superhabitability.

The *Atmos* model consists of the two main modules: climate and photochemistry. The climate module computes the 1D temperature profile of the planet's atmosphere across different altitudes (pressure levels) (Kasting and Ackerman 1986). Concurrently, the photochemistry module computes the concentrations of numerous gases (here: species), including $O_2$, $O_3$, $CO_2$, $H_2O$, $CH_4$, and $N_2O$, among others, as a function of altitude (pressure levels) (Zahnle *et al.* 2006). The two modules alternate to achieve consistent solutions for chemical and temperature profiles, essential for steady-state atmospheric conditions over extended periods. The photochemistry module contains 61 chemical species involved in over 300 chemical reactions typical of Earth's atmosphere. The *Atmos* model can be run either uncoupled or coupled. When coupled, the climate module's temperature profile impacts the abundance of gases and aerosols predicted by the photochemistry module, and vice versa, allowing for the interplay of greenhouse gases and temperature to be consistently modeled.

For the climate module in *Atmos*, absorption coefficients are derived from the HITRAN 2008 and HITEMP 2010 databases[1], covering pressures from $10^{-5}$ to $10^2$ bar and temperatures from 100 to 600 K. *Atmos* uses the correlated-k method for absorption calculations and solves radiative transfer using the delta two-stream approximation (Toon *et al.* 1989).

The water vapor profiles were calculated in the climate module using the methodology of Manabe and Wetherald (1967), which assumes a constant relative humidity throughout the atmosphere. Here, relative humidity is defined as the ratio of the partial pressure of water vapor to the saturation vapor pressure, which is temperature-dependent and calculated via the Clausius-Clapeyron equation. This method allows for a consistent water vapor distribution that adjusts realistically to atmospheric temperature and pressure changes (see **Supplementary Materials** for detailed derivations).

The photochemistry module allows the user to choose the abundance of atmospheric species based on different upper ($M_{BOUND}$) and lower ($L_{BOUND}$) boundary conditions. The upper boundary conditions describe the input and loss of species between the uppermost layer of the planet's atmosphere and space, whereas the lower boundary conditions determine the interactions of atmospheric species with the planetary surface. In this work we only modified the lower boundary conditions, which include:

---

[1] The HITRAN (High-Resolution Transmission) database provides detailed information about the absorption of light by atmospheric gases, while the HITEMP database offers similar information but for high temperatures.

- $L_{BOUND} = 0$ (Deposition Velocity, $V_{dep}$) – Represents the deposition of species onto or at the surface, accounting for reactions with surface rocks, dissolution in water, or consumption by biological processes.
- $L_{BOUND} = 1$ (Fixed Mixing Ratio, Fixed$_{MR}$) – Fixes the mixing ratio at the surface layer and determines the required gas flux to maintain that concentration. This is often used for gases like oxygen, where the fixed mixing ratio is set to 21% for modern Earth.
- $L_{BOUND} = 2$ (Constant Upward Flux, SG$_{FLUX}$) – Similar to $V_{dep}$ but in the opposite direction, representing the flow rate of gases from the surface into the atmosphere in molecules/cm²/s.
- $L_{BOUND} = 3$ (Volcanoes) – SG$_{FLUX}$ is distributed equally over a specific number of atmospheric layers (DIST$_H$). For example, if SG$_{FLUX}$ is 10 and DIST$_H$ is 10 km, each kilometer has an SG$_{FLUX}$ value of 1. Some flux is then removed by $V_{dep}$.

### 2.1.1. Atmos input spectra and simulation workflow

In this work we simulated three hypothetical superhabitable planets around stars with effective temperatures of 3900K, 4300K and 4900K, respectively. For comparison, we also simulated modern Earth orbiting the Sun. To connect our study with an existing terrestrial planet orbiting a K dwarf star, we simulated the conditions on Kepler-62e, which may be either a potentially rocky, water-covered planet or a mini-Neptune. With a favorable equilibrium temperature of 270 K, an incident stellar flux of 1.2 S₀, and an orbital period of 122 days, Kepler-62e lies within the habitable zone, suggesting it could support life if it is rocky and adequate cloud cover is present (see **Section 2.1.2.** and **Supplementary Materials**) (Borucki *et al.* 2013, Kaltenegger *et al.* 2013). In this study, we only consider the rocky planet possibility.

In this work, we used model stellar spectra from the PHOENIX spectral library (see **Figure 1**) (Husser *et al.* 2013). For the solar spectrum we used a default composite spectrum from *Atmos*. Metallicity Z denotes the total mass fraction of all elements heavier than hydrogen and helium in a star, usually expressed as a logarithmic (base 10) measure of the star's metallicity relative to the Sun. We selected a near-solar metallicity (Z = 0.0) for the stars with temperatures of 3900K, 4300K, and 4900K, while we chose a sub-solar metallicity (Z = –0.5; corresponding to [Fe/H] ~ –0.32) for the Kepler-62 PHOENIX star to most closely match its calculated metallicity of [Fe/H] = –0.37 ±0.04 (Kepler-62 2024). This indicates that the metal content of the Kepler-62-like star in our study is 10⁻⁰·⁵ or roughly 31.6% of that of the Sun.

**Figure 1:** Stellar input spectra for the 1D coupled climate-photochemistry model *Atmos*. The spectra are from the PHOENIX spectral library. For the solar spectrum, the default composite spectrum in *Atmos* is used. Z represents the total mass fraction of all elements heavier than hydrogen and helium in a star, and typically refers to the logarithmic (base 10) measure of the star's metallicity relative to the Sun.

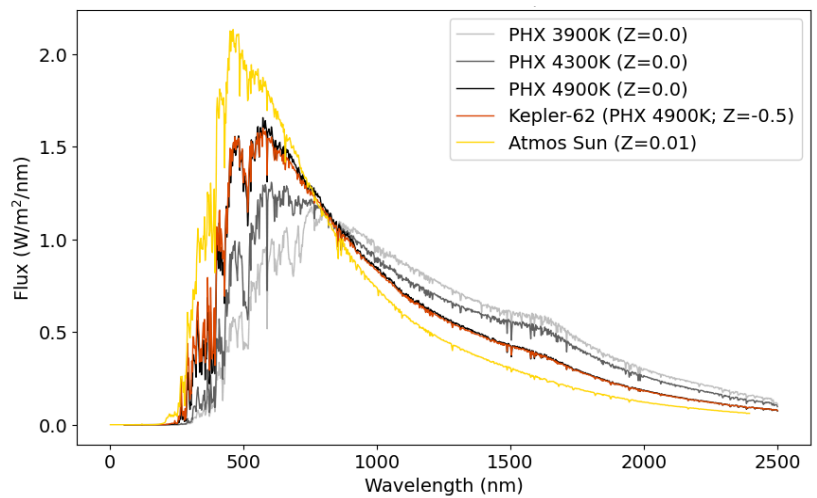

The PHOENIX model spectra lack an accurate representation in the ultraviolet (UV; 100–400 nm) range, particularly the far UV (FUV) and near UV (NUV) fluxes, which are crucial for photochemistry. This radiation plays a role in initiating the photolysis of water ($H_2O$), which generates hydroxyl (OH) radicals. These OH radicals participate in catalytic cycles that can influence the concentrations of several critical compounds in the atmosphere (Segura *et al.* 2005, Hu *et al.* 2012, Arney *et al.* 2017, Rugheimer and Kaltenegger 2018). Nevertheless, these spectra offer a baseline assumption–the very small UV flux emerging from the black body spectrum–, and even minor variations in this baseline can significantly alter abundance profiles. Atmospheric temperature differences also drive the photochemical module as the gas phase reactions have a temperature dependence. Furthermore, the UV contribution to the overall spectral energy distribution of the stars is relatively low, and the stellar bands in the climate module do not even extend down to FUV. The specific UV wavelengths and distribution of lines that are critical for photochemistry have minimal impact on the results of the climate module. Therefore, this work did not emphasize intensive photochemistry modeling with full coupling to stellar spectra that are unsuitable for photochemistry modeling.

To assess the impact of UV flux on the transmission spectrum of a planetary atmosphere, we conducted a sensitivity study by evaluating the influence of a UV-active star spectrum (using HD 85512, which includes full FUV and NUV fluxes) compared to the PHOENIX spectrum of a 4300K star with only baseline UV. This comparison allowed us to explore how a full UV radiation profile impacts the resulting transmission spectrum by altering atmospheric composition and photochemical pathways. The study showed that percent differences in transmission spectra are small to moderate, with values of 0.21% ± 0.14% for the absolute transmission spectrum and 11.45% ± 11.03% for the differential transmission spectrum. Here, the absolute spectrum represents the direct transit depth values in parts per million (ppm), while the differential spectrum shows changes in transit depth relative to a baseline value, enhancing sensitivity to smaller variations. The UV-active scenario is discussed in detail in the discussion **Section 4.2.** and **Supplementary Materials (Fig. S8)**.

Instead, we conducted one iteration of the photochemistry module until convergence was achieved (i.e., until the time steps approached the age of the Universe) to establish all parameters for the climate module. Following this, we ran the climate module for one iteration until convergence was reached, specifically when the Divergence from Root Mean Square (DIVFrms) – a measure of the variability or deviation of a set of values from their root mean square (RMS) – reached an order of magnitude of $10^{-3}$. Finally, we performed one last iteration of the photochemistry module for plotting and visualization purposes, following the sequence: Photochemistry → Climate → Photochemistry. For the scenarios in this study, we simplified our approach by assuming a lower boundary condition with a constant volume mixing ratio (vmr; $L_{BOUND}$ = 1) for our simulation sequence. The vmr is defined as the ratio of the number of molecules of a particular gas to the total number of molecules of all gases present in the atmosphere. This was done instead of assigning surface fluxes, which would require a more accurate representation in the UV range. Volcanic outgassing fluxes were kept constant and are not influenced by biological activity.

### *2.1.2. Atmos scenarios*

#### *Superhabitable planets*

In this work, we simulated three pairs of hypothetical superhabitable planets, each pair corresponding to two locations within the HZs of K dwarf host stars with temperatures of 3900K, 4300K, and 4900K. The radii and masses of the host stars were derived from stellar evolution tracks for stars that are 5 billion years old (Baraffe *et al.* 2015). The stellar radii (R) are 0.54, 0.62 and 0.74 solar radii ($R_\odot$), respectively. The corresponding stellar masses are 0.57, 0.67, and 0.80 solar masses ($M_\odot$), respectively.

At the two locations within the HZs, the planets receive stellar fluxes (F) of 0.6 and 0.8 times the solar constant ($S_0 = 1366$ W/m$^2$) at the top of their atmospheres. These flux levels roughly correspond to the centers of the HZs and the midpoint between the inner edge and the centers of the HZs, respectively (Kopparapu *et al.* 2013). For detailed calculations of stellar luminosities, star-planet distances (i.e., semi-major axes), planetary masses, surface gravities and orbital periods, refer to the **Supplementary Materials**. The semi-major axes for the superhabitable planet in orbit in the HZ around a 3900K host star coincides with a range of distances where the planet is tidally locked (Kasting *et al.* 1993). **To provide more living space and support a more energetic biosphere while maintaining the essential processes of plate tectonics, we designed the SH planets in this study to be 25% larger in radius than Earth. A full summary of stellar and planetary parameters used in this work can be seen in Table 1 (a).** To estimate the mass of a 1.25 Earth-radius planet, we assumed Earth-like density, resulting in an approximate mass of 1.95 Earth masses. Although gravitational compression due to increased mass in larger rocky planets typically increases density (Chen and Kipping 2016), using Earth's density provides a practical and consistent baseline, given that specific density profiles for superhabitable planets are speculative. While a more accurate empirical scaling would suggest a mass closer to 2.2 Earth masses (Chen and Kipping 2016), this difference only moderately increases surface gravity (from 12.25 m/s$^2$ to 13.79 m/s$^2$) and reduces atmospheric scale height (by ~11%$^2$), minimally impacting transit depth.

In terms of the composition of a superhabitable atmosphere, we focused on key organisms and biological sources influencing the Earth's biosphere and their atmospheric fingerprints. These include $O_2$, $H_2$, $CH_4$, $N_2O$ and $CO_2$. The summary of their corresponding sources and processes is shown in **Table 2**. As mentioned, we assigned constant mixing ratios to each of these species. For $CH_4$ and $N_2O$, their values were increased to ten times the present atmospheric levels (PAL) to simulate the increased production by methanogenic microbes and (de)nitrifying bacteria, respectively. The mixing ratio of $H_2$ was decreased to 0.1 PAL to simulate the increased consumption by enzymes. The mixing ratio of $O_2$ was increased from present-day 21% by volume on Earth to 25% to simulate a photosynthetically thriving biosphere. Moreover, we increased the surface RH from 80 to 90% for SH planets and Kepler-62e, since higher RH significantly influences biodiversity and water acts as a universal solvent, essential for dissolving and transporting substances essential to biophysical processes and ecosystem productivity (Deng *et al.* 2016).

Due to the significant temperature drop in 1D models when incident stellar flux decreases, we adjusted the $CO_2$ mixing ratio to maintain a surface temperature of ~294 K. We conducted a parameter study to examine the effects of varying surface pressures (1.5 to 3 bar) and $CO_2$ levels (10 to 1800 PAL) on surface temperatures, focusing on a superhabitable planet orbiting a 4300K star with 0.6 and 0.8 $S_0$. We considered a temperate range of 288–298 K for superhabitable planets, with the upper limit, 298 K, corresponding to conditions during the Carboniferous Era on Earth, a period known for high biomass and biodiversity (Vilović *et al.* 2023). In the 0.6 $S_0$ scenario, achieving 294 K requires at least 1800 PAL of $CO_2$ at 1.5 bar, which is the maximum the atmosphere can hold without exceeding total pressure. Similarly, at 2.0 bar, temperatures above 295 K are unattainable. Thus, a surface pressure of 2 bar was selected for the six superhabitable scenarios (three K dwarf stars with two localizations in the HZ each) to maintain Earth-like conditions while compensating for reduced stellar flux. At this pressure, 1500 PAL (60% by volume) of $CO_2$ is required to sustain ~294 K at 0.6 $S_0$, whereas only 80 PAL (3% by volume) is sufficient at 0.8 $S_0$ (see **Table 1 (b)** and **Figure 2**). While these elevated $CO_2$ levels correspond more closely to the anoxic conditions of the Archean Eon on Earth (Lehmer *et al.* 2020) than to levels following the Cambrian explosion that diversified Earth's biosphere, we included this scenario to examine the range of atmospheric conditions required to sustain habitable temperatures under reduced stellar flux. For the 0.6 $S_0$ planets, the term 'superhabitable' should be interpreted conservatively, as these scenarios diverge from typical Earth-like conditions. Nonetheless, it provides valuable insights into the thermal requirements for planets located near the center of the habitable zone, which may offer enhanced resistance against moist or runaway greenhouse states at the inner edge of the HZ, as proposed by (Heller and Armstrong 2014). Similarly, because Kepler-62e receives 1.2 $S_0$, we reduced $CO_2$ concentrations to 0.1 PAL to avoid additional surface warming, which results in a surface temperature of 313 K.

---

$^2$ Assuming a planetary surface temperature T=294K and Earth's mean molecular weight μ = 0.02897 kg/mol, the scale height with surface gravity of 12.25 m/s$^2$ = 1.14x10$^{-20}$ meters, and the scale height with surface gravity of 13.79 m/s$^2$ = 1.02x10$^{-20}$ meters; resulting in ~11% scale height difference.

**Table 1.** Stellar/planetary parameters and atmospheric compositions used in the coupled climate-photochemistry model *Atmos*. (a) Input parameters for the modern Earth, superhabitable (SH) planet scenarios, and Kepler-62e. Two positions within the habitable zone are considered for the SH scenarios, with stellar fluxes of 0.6 and 0.8 times the solar constant ($S_0$ = 1366 W/m$^2$). (b) Atmospheric composition used in the model with comprehensive lower (surface) boundary conditions for most dominant and biologically relevant chemical species. The boundary condition parameters include a constant deposition velocity (Vdep; Lbound=0), a constant mixing ratio (Fixed MR; Lbound=1), a constant upward flux (SG Flux; Lbound=2) and a constant deposition velocity with a vertically distributed upward flux (Vdep + SG Flux + Dist H, where Dist H is the number of atmospheric layers; Lbound=3). Here we simplify our approach by only assigning constant volume mixing ratios (Lbound = 1) for our simulation sequence.

**(a)**

| | | Earth-Sun System | Superhabitable 3900K System '0.6'/'0.8' | Superhabitable 4300K System '0.6'/'0.8' | Superhabitable 4900K System '0.6'/'0.8' | Kepler-62e System* |
|---|---|---|---|---|---|---|
| Properties of host star | Effective temperature [K] | 5772 | 3900 | 4300 | 4900 | **4925** |
| | Mass [M$_\odot$] | 1 | 0.57 | 0.67 | 0.80 | **0.69** |
| | Radius [R$_\odot$] | 1 | 0.54 | 0.62 | 0.74 | **0.64** |
| | Metallicity [Z] | 0.01 | 0 | 0 | 0 | **−0.5** |
| | Age [Gyr] | 4.5 | 5 | 5 | 5 | **~7** |
| Planetary properties | Incident stellar flux [S$_0$] | 1 | **0.6 / 0.8** | **0.6 / 0.8** | **0.6 / 0.8** | **1.2** |
| | Semi-major axis [AU] | 1 | **0.32 / 0.28** | **0.44 / 0.38** | **0.69 / 0.59** | **0.43** |
| | Orbital period [days] | 365 | **88 / 72** | **130 / 105** | **234 / 185** | **122** |
| | Gravitational acceleration [cm/s$^2$] | 980 | 1225 | 1225 | 1225 | **1225** |
| | Radius [R$_\oplus$] | 1 | 1.25 | 1.25 | 1.25 | **1.61** |
| | Mass [M$_\oplus$] | 1 | 1.95 | 1.95 | 1.95 | **~3.3** |
| | Density [ρ$_\oplus$] | 1 | 1 | 1 | 1 | **0.78** |
| | Surface pressure [P$_\oplus$] | 1 | 2 | 2 | 2 | **~1.56** |
| | Surface temperature [K] | 288 | 294 | 294 | 294 | **~313** |
| | Surface relative humidity | 80% | 90% | 90% | 90% | **90%** |

**(b)**

| | **Modern Earth** | | | | | **Superhabitable planets '0.6'/'0.8' and Kepler-62e** | | | | |
|---|---|---|---|---|---|---|---|---|---|---|
| Chemical species | Boundary condition (Lbound) | Deposition Velocity [cm/s] | Constant Mixing Ratio | Surface Flux [molecules/ cm²/s] | Layers (Dist H) | Boundary condition (Lbound) | Deposition Velocity [cm/s] | Constant Mixing Ratio | Surface Flux [molecules/ cm²/s] | Layers (Dist H) |
| $N_2$ | 1 | - | 78% | - | - | 1 | - | fill gas | - | - |
| $O_2$ | 1 | - | 21% | - | - | 1 | - | 25% | - | - |
| $H_2$ | 0 | $2.46 \times 10^{-4}$ | $5.30 \times 10^{-7}$ | - | - | 1 | - | 0.1 PAL | - | - |
| $CH_4$ | 2 | - | $1.80 \times 10^{-6}$ | $1.00 \times 10^{11}$ | - | 1 | - | 10 PAL | - | - |
| $N_2O$ | 2 | - | $3.00 \times 10^{-7}$ | $1.53 \times 10^{9}$ | - | 1 | - | 10 PAL | - | - |
| $CO_2$ | 3 | $5.00 \times 10^{-5}$ | $4.00 \times 10^{-4}$ | $6.88 \times 10^{8}$ | 14 | 1 | - | **1500/80/0.1 PAL | - | - |

*Stellar and planetary parameter data from (Borucki *et al.* 2013). Assumptions were made for the gravitational acceleration and the surface relative humidity.

**To maintain a temperate surface, 1500 PAL of $CO_2$ is used for the 0.6 $S_0$, while 80 PAL of $CO_2$ is used for the 0.8 $S_0$ scenario. 0.1 PAL of $CO_2$ is used for Kepler-62e.

**Table 2.** Key sources influencing the Earth's biosphere and their atmospheric fingerprints.

| Source/Organism | Signature |
|---|---|
| **Cyanobacteria, plants, green algae** | **$O_2$:** These organisms are responsible for a significant flux of molecular oxygen into the Earth's atmosphere. Cyanobacteria laid the foundation for oxygenic photosynthesis, which was later adopted by plants and algae (Schirrmeister *et al.* 2015). In a superhabitable atmosphere, molecular oxygen levels could increase from present-day 21% by volume on Earth to 25%, reflecting a thriving photosynthetic biosphere that enhances overall biospheric productivity. |
| **Hydrogenases enzymes** | **$H_2$:** Molecular hydrogen is consumed by hydrogenases – specialized metalloenzymes that catalyze the reversible oxidation of molecular hydrogen into protons and electrons. This process is crucial in microbial energy metabolism (Lane *et al.* 2010, Greening and Boyd 2020). In a superhabitable atmosphere, $H_2$ levels could be decreased tenfold to reflect this consumption. |
| **Methanogenic microbes** | **$CH_4$:** Methane is released into the atmosphere via methanogenesis – a process by which these anaerobic archaea produce methane as a metabolic byproduct (Wen *et al.* 2017). In a superhabitable atmosphere, methane levels could be increased tenfold. |
| **Fungi and (de)nitrifying bacteria** | **$N_2O$:** Fungi are vital members of our biosphere because they break down organic matter and release carbon, oxygen, phosphorus and nitrogen into the soil and atmosphere (Tansey and Brock 1972). Bacteria, such as *Pseudomonas*, *Paracoccus*, and *Bacillus*, release nitrogen into the atmosphere via denitrification (Averill and Tiedje 1982). In a superhabitable atmosphere, nitrous oxide levels could be up to ten times higher. |
| **Respiration by aerobic organisms, decomposition of dead matter and volcanic activity** | **$CO_2$:** The main natural source of carbon dioxide production is the respiration of living organisms, followed by decomposition of organic matter (Grace 2004, Battin *et al.* 2009). Carbon is necessary for the formation of complex molecules essential for life such as DNA (Schulze-Makuch and Irwin 2006) and is one of our planet's primary greenhouse gases. Volcanic activity and ocean-atmosphere exchange also play important roles. In a superhabitable atmosphere, these levels would be regulated by a balance between biological productivity and greenhouse effects, requiring between 3% and 60% by volume to offset reduced stellar K dwarf fluxes compared to our Sun. |

**Figure 2**: Parameter study on the effect of varying planetary surface pressures and $CO_2$ levels on surface temperatures of a superhabitable planet orbiting a 4300K star. Two incident stellar flux scenarios within the habitable zone are shown: 0.6 (solid lines) and 0.8 (dashed lines) solar constants ($S_0 = 1366$ W/m²). The shaded green area represents the temperate range of 288–298 K, with 298 K corresponding to the Carboniferous Era on Earth, a period of high biomass and biodiversity. In the 0.6 $S_0$ scenario at 1.5 bar, 294 K cannot be achieved with less than 1800 PAL of $CO_2$ due to atmospheric pressure limits.

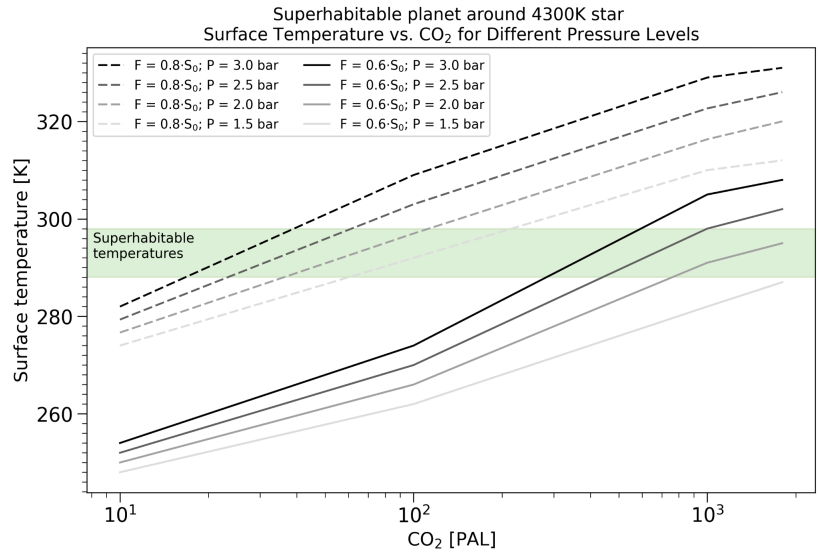

## Modern Earth around the Sun

For simulating modern Earth around the Sun, we used the 1976 US Standard Atmosphere, which extends from 0 to 80 km in altitude (~$10^{-5}$ bar). The simulation includes modern-day greenhouse gases: 400 ppmv of $CO_2$, an upward surface $CH_4$ flux of $1.00 \times 10^{11}$ molecules/cm²/s and an upward surface $N_2O$ flux of $1.53 \times 10^9$ molecules/cm²/s. Additionally, a water vapor ($H_2O$) profile was calculated using the Manabe and Wetherald method. The concentrations of $O_2$ and $O_3$ were also set to present-day levels (see **Table 1 (b)**).

## Kepler-62e

Based on SH criteria, Kepler-62e emerges as a promising target worth exploring further. Most of the known parameters for this planet are adopted from (Borucki *et al.* 2013). Kepler-62e is about 990 light-years (300 parsecs) from Earth. The planet has an equilibrium temperature of 270 K, an incident stellar flux of 1.2 $S_0$, a planetary radius of 1.61 $R_\oplus$ as well as an upper limit on surface pressure of 1.56 $P_\oplus$. If it has sufficient cloud cover to reflect radiation, Kepler-62e might sustain surface water, and it lies within the habitable zone (Borucki *et al.* 2013). It has further been suggested that Kepler-62e could be a water planet with a mass between 2–4 $M_\oplus$ (Kaltenegger *et al.* 2013). Kepler-62e scores 0.83 on the Earth Similarity Index (ESI) and 0.87 on the Statistical-likelihood Exo-Planetary Habitability Index (SEPHI) (Schulze-Makuch *et al.* 2011, Rodríguez-Mozos and Moya 2017). These factors make Kepler-62e particularly interesting for the potential development of life. We assumed a gravitational acceleration of 1225 cm/s², to match those of the SH planets.

For the purpose of this study, we assume that Kepler-62e is a rocky planet with both land and ocean regions, rather than a pure water world. This assumption mitigates the challenges associated with nutrient transport in deep oceans, which could limit biological productivity and complex life (Holland 2006, Kite and Ford 2018). The presence of land enhances habitability by providing diverse ecological niches, greater nutrient availability, and direct atmospheric contributions through photosynthetic and geothermal processes. This mixed-landscape assumption improves the likelihood of detecting biosignature gases like oxygen, methane, and nitrous oxide in the atmosphere, as land regions can release these gases directly into the atmosphere, enhancing their detectability (Schwieterman *et al.* 2018). We also assume a superhabitable atmospheric composition, with the exception of $CO_2$, which we reduce tenfold from the modern Earth value to offset Kepler-62e's increased stellar flux (**Table 1 (b)**). For detailed calculations of the density and corresponding mass of Kepler-62e, refer to the **Supplementary Materials**. The full summary of stellar and planetary parameters for the Kepler-62 system can be found in **Table 1**.

## *2.2. POSEIDON*

We used the *POSEIDON* exoplanet atmosphere model to calculate synthetic planetary spectra based on the T–P profiles and atmospheric outputs from *Atmos* (MacDonald and Madhusudhan 2017, MacDonald 2023). The code is designed for calculating the transmission spectrum of an exoplanet during its transit of the host star. It simulates the day-night boundary of the atmosphere, assuming hydrostatic equilibrium and an average temperature structure and chemistry at the terminator. The full geometry of the line-by-line radiative transfer computation can be discerned from Figure 1 in (MacDonald and Madhusudhan 2017). The model atmosphere is divided into 200 evenly spaced layers in the pressure regimes between $10^{-11}$ and 2 bar for the SH planet scenarios, between $10^{-9}$ and 1.56 bar for Kepler-62e, and between $10^{-8}$ and 1 bar for modern Earth around the Sun. We provide the host star and planet properties used for *Atmos* in **Table 1**. We assumed $O_2$, $CO_2$, $CH_4$, $H_2O$, $O_3$ and $N_2O$ as the major chemical species in the model atmosphere, as these species are spectrally active in the wavelength regime between 0.3 and 14 micrometers, mostly covered by JWST. *POSEIDON* offers a specialized 'temperate' opacity database for terrestrial planets, tailored for low-temperature atmospheres (below 400 K) using HITRAN line lists. After the atmospheres were created, *POSEIDON* computed the corresponding transmission spectra by means of radiative transfer.

## *2.3. PandExo*

To simulate observations of transiting exoplanets with the JWST, we utilized the *PandExo* tool (Batalha *et al.* 2017). *PandExo* is available both as an online tool and a Python package for generating instrument simulations of JWST's NIRSpec, NIRCam, and NIRISS, as well as MIRI LRS. Here we focused on the wavelength range from ~1 to 12 micrometers covered by the high-resolution NIRSpec G140H (0.97 − 1.89 µm), -G235H (1.66 − 3.17 µm), -G395H (2.87 − 5.27 µm) and the low-resolution MIRI LRS (5.0 − 12.0 micrometers) instruments. The high-resolution NIRSpec instruments include the following band gaps, which partition the data into the 'NRS1' and 'NRS2' detectors:

- NIRSpec G140H: Band gap at 1.335 µm
- NIRSpec G235H: Band gap at 2.235 µm
- NIRSpec G395H: Band gap at 3.77 µm

The input parameters for *PandExo* (stellar and planetary radii as well as stellar effective temperatures) are consistent with those used in *Atmos* and *POSEIDON* (**Table 1 (a)**). Here the general steps included:

1. Generating simulated JWST observations around a fiducial flat spectrum with a full resolution (R~3000) using *PandExo*.
2. Scattering the simulated observations about a *POSEIDON* forward model; and
3. Running an atmospheric retrieval on the synthetic JWST observations.

We placed the six superhabitable planets as well as Kepler-62e and modern Earth around the Sun at a distance of 30 parsecs from us for better comparison. The calculations of the corresponding apparent J-band magnitudes at this distance can be found in the **Supplementary Materials**. They are 8.39, 6.89, 6.39, 6.39 and 6.0 for stars of temperatures 3900K, 4300K, 4900K, Kepler-62 and the Sun, respectively. These calculated brightnesses of the host stars fall within the magnitude range covered by the selected instruments. The signal-to-noise ratio (S/N) for an observation is proportional to $\sqrt{N_{trans}}$, with $N_{trans}$ denoting the number of transits.

Our ultimate objective was to determine the key characteristics of an exoplanet's atmosphere—such as the T–P profile and chemical abundances—based on the observed transmission spectra. We accomplished this by running a retrieval on the simulated JWST data for a specific number of transits using the PyMultiNest algorithm with a resolution of R=10,000. For this process, we defined the true values of the parameters. For vertical volume mixing ratio profiles of the atmospheric species $O_2$, $CO_2$, $CH_4$, $H_2O$, $O_3$ and $N_2O$, the true value is typically taken at a pressure close to the maximum of the contribution function, around $10^{-2}$ bar for transmission spectra. Therefore, we used the log of the volume mixing ratios at 10 mbar as our 'true' values. The posterior distributions and joint histograms we obtained through the retrieval process give us an idea of how well-constrained specific atmospheric species are for a specific number of transits. For example, narrow and circular distributions imply well-constrained, uncorrelated parameters. Elliptical joint distributions indicate well-constrained but correlated parameters, whereas wide and diffuse distributions suggest poorly constrained parameters, whether correlated or not.

## 3. Results

We used the *Atmos* model to define the SH parameter space, producing T–P and atmospheric abundance profiles for key biomarkers. We utilized *POSEIDON* to calculate synthetic planetary transmission spectra from these profiles. Finally, we used the *PandExo* tool to simulate JWST observations and performed atmospheric retrievals to assess the feasibility of the lowest number of transits required to constrain key spectral features indicative of life on Earth.

### 3.1. Temperature Profiles

**Figure 3** depicts the T–P profiles for various SH planet scenarios within the HZ of K dwarf host stars with temperatures of 3900K, 4300K, and 4900K. The detailed planetary and stellar parameters, along with the atmospheric boundary conditions, are provided in **Table 1**. These profiles are shown for two levels of stellar flux: 0.6 and 0.8 times the solar constant ($S_0$ = 1366 W/m²), represented by solid and dashed lines, respectively. These flux levels correspond to the center of the HZs and the midpoint between the inner edge and the center of the HZs, respectively. For comparison, the T–P profiles of modern Earth, with a surface temperature of 288.4 K at 1 bar, and Kepler-62e, with a surface temperature of 313.6 K at 1.56 bar, are shown. SH planets with surface pressures of 2 bar, orbiting stars at 3900K, 4300K, and 4900K, show surface temperatures around 294.4 K, with a standard deviation of ±0.3 K. All scenarios show a typical temperature decrease from the surface up to around $10^{-1}$ bar, resulting from the imposed adiabatic lapse rate.

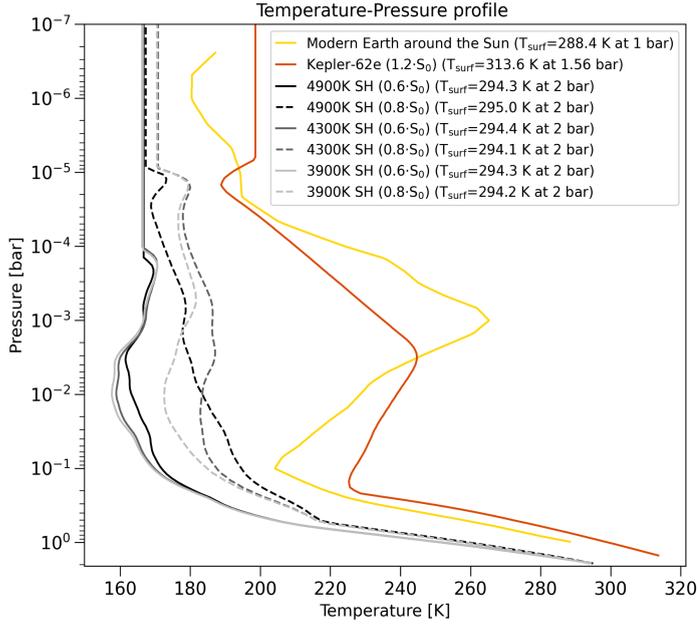

**Figure 3:** Temperature vs. pressure profiles of various superhabitable (SH) planet scenarios at two locations within the habitable zones (HZ) of K dwarf host stars with temperatures of 3900K, 4300K, and 4900K. At the two locations within the HZs, the planets receive stellar fluxes of 0.6 and 0.8 times the solar constant ($S_0 = 1366$ W/m$^2$) at the top of their atmospheres (solid and dashed lines, respectively). The modern Earth around the Sun is also plotted for comparison. To bridge this study to an existing terrestrial planet around a K dwarf star, we also simulate Kepler-62e. The planetary and stellar parameters as well as the atmospheric boundary conditions are shown in **Table 1**.

### 3.2. Abundance Profiles

**Figure 4** shows the volume mixing ratio profiles of selected biomarkers–ozone ($O_3$), water vapor ($H_2O$), carbon dioxide ($CO_2$) and methane ($CH_4$)–for the same scenarios as in **Figure 3**. The $O_3$ profiles exhibit similar trends throughout all scenarios: a gradual increase in vmrs with decreasing pressures, followed by a moderate inversion around $10^{-1}$–$10^{-2}$ bar, and a secondary inversion at around $10^{-5}$ bar. $O_3$ concentrations decrease with decreasing UV environments, with a minimum for the SH planets around the 3900 K host star. The $H_2O$ profiles all exhibit a linear decrease in concentrations up to around $10^{-1}$ bar, followed by a slight increase for the SH scenarios, and isoconcentration profiles for Kepler-62e and modern Earth around the Sun. We note that photochemical effects are not included in the model (as discussed in **Section 2.1.1.**), which likely leads to an overestimate of transit features for $H_2O$ and $CH_4$. The near-constant abundances of $CH_4$ and $CO_2$ are consistent with the input concentrations for the respective scenarios. For Kepler-62e, the concentration of $CH_4$ decreases to a constant value between $10^{-1}$ and $10^{-2}$ bar. In contrast, for modern Earth around the Sun, $CH_4$ concentrations continue to decrease beyond this pressure range, dropping by a couple of orders of magnitude.

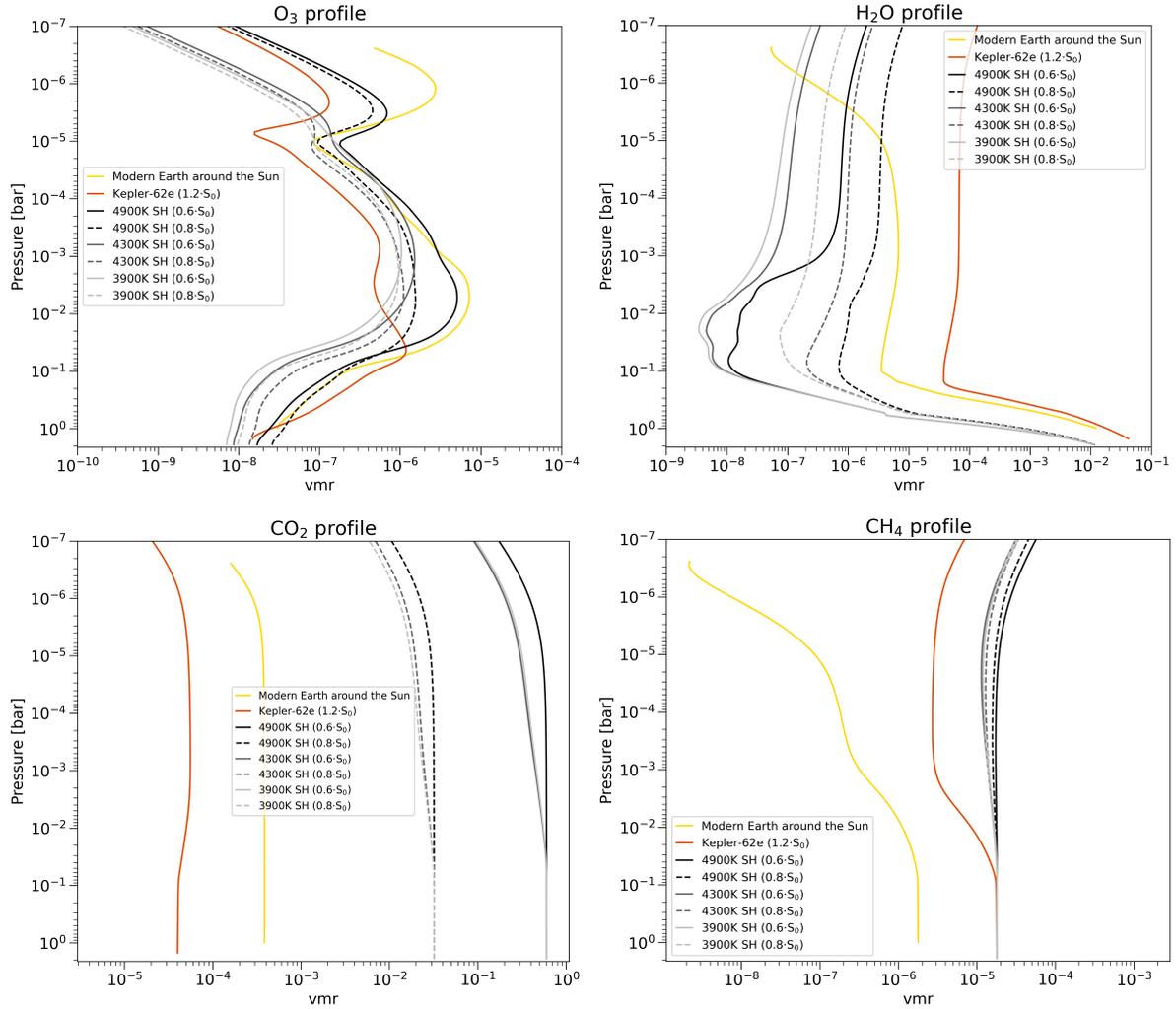

**Figure 4:** Atmospheric volume mixing ratio (vmr) abundance profiles of selected biomarkers–ozone ($O_3$), water vapor ($H_2O$), carbon dioxide ($CO_2$) and methane ($CH_4$)–for various superhabitable (SH) planet scenarios at two locations within the habitable zones (HZ) of K dwarf host stars with temperatures of 3900K, 4300K, and 4900K. Line descriptions are consistent with those in **Figure 3**.

### 3.3. Transmission Spectra

The transmission spectra for various superhabitable planet scenarios as well as Kepler-62e and modern Earth around the Sun, calculated using the *POSEIDON* model are shown in **Figure 5**. Since the cross-sectional area of a sphere, such as a planet or star, is given by $\pi R^2$, where R is the radius, the fraction of the star's light blocked during a transit (transit depth) is proportional to the square of the ratio of the planet's radius to the star's radius $(R_p/R_\star)^2$. Kepler-62e shows the highest transit depth compared to the SH planets and Earth around the Sun, consistent with its large radius relative to the size of the host star. Modern Earth around the Sun, being the smallest planet with the largest stellar radius in comparison, exhibits the lowest transit depth. The SH planets display varying transit depths based on their host star temperature and incident stellar flux. Generally, the 0.8 $S_0$ scenarios (black, dark, and light gray solid lines) show slightly higher base transit depths and more prominent spectral features than their 0.6 $S_0$ counterparts (dashed lines).

**Figure 5:** Transmission spectra of various superhabitable (SH) planet scenarios as well as Kepler-62e and modern Earth around the Sun. Line descriptions are consistent with those in **Figure 3**. The absolute transit depth is the fractional decrease in the star's brightness as the planet passes in front of it (transits). It is a measure of the planet's cross-sectional area relative to the star's cross-sectional area by squaring the ratio of the planetary radius $R_p$ to the stellar radius $R_*$ expressed in parts per million (absolute transit depth times $10^6$).

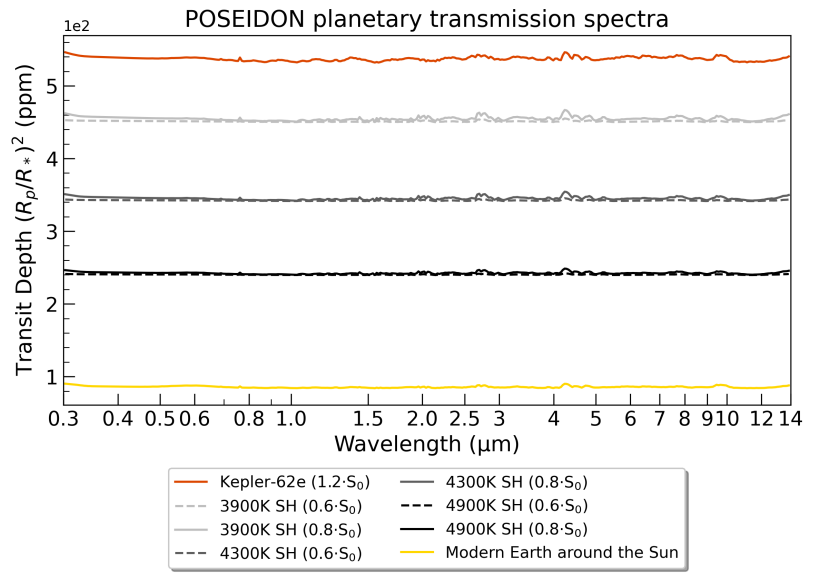

**Figure 6** shows the transmission spectra from **Figure 5** with the transit depths expressed as differences from a baseline transit depth. This highlights the relative amplitudes of individual spectral features across the scenarios. The key absorption features of major atmospheric constituents in the *POSEIDON* wavelength range of 0.3–14 µm are annotated in the figure and summarized in **Table 3**. This comparison highlights differences in individual spectral features and illustrates the combined effect of changes in spectral feature amplitudes and absolute transit depths across different scenarios. The transmission spectra highlight key spectral features, including several biosignature bands: the $O_2$ band at 0.76 µm, $N_2O$ bands in the near infrared (NIR), the $O_3$ Hartley-Huggins band between 0.31 and 0.34 µm, the $O_3$ Chappuis band spanning 0.4 to 0.85 µm, and the $O_3$ fundamental band near 9.6 µm. Additionally, the spectra show prominent $CO_2$ absorption bands around 4.3 µm, numerous rotational-vibrational bands of $CH_4$ and $H_2O$ in the NIR and infrared (IR) regions, and prominent Rayleigh scattering in the UV and visible light spectrum below 1 µm (which appears as a steep increase in transit depth at shorter wavelengths). For the 0.6 $S_0$ SH scenarios (upper panel), most individual spectral features have smaller amplitudes than modern Earth around the Sun, except for the $CO_2$ features at 1.6 and 2.0 µm, and they do not exhibit a prominent Rayleigh slope. Kepler-62e, on the other hand, shows the highest amplitudes for all spectral features and an even steeper Rayleigh slope. Contrastingly, the 0.8 $S_0$ SH scenarios exhibit higher spectral feature amplitudes than modern Earth around the Sun, with the 2.0, 2.7 and 4.3 µm $CO_2$ features also surpassing those of Kepler-62e.

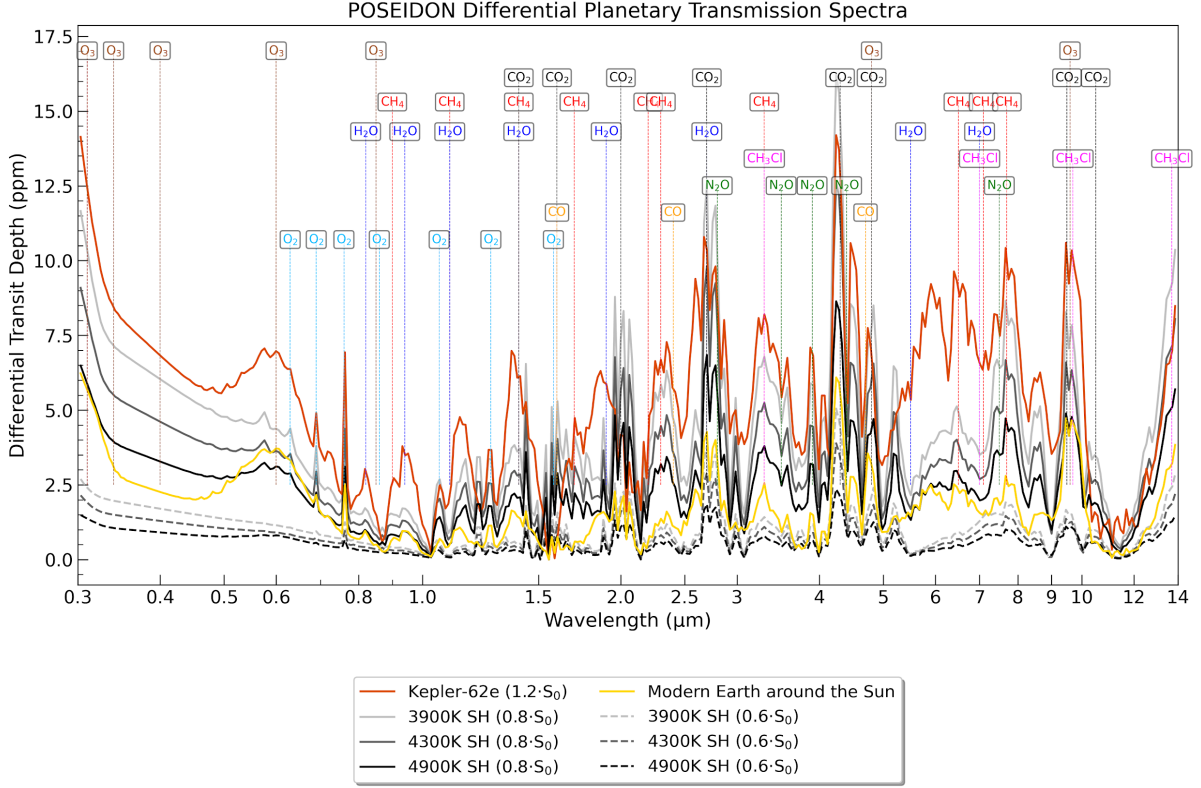

**Figure 6:** Differential transmission spectra of various superhabitable (SH) planet scenarios, as well as Kepler-62e and modern Earth around the Sun, calculated using the forward model code *POSEIDON*. Spectral features are displayed in parts per million (ppm) relative to each planet's baseline transit depth, enhancing the visibility of individual molecular signatures. Key absorption features of major atmospheric constituents, including $O_2$, $O_3$, $CH_4$, $H_2O$, $N_2O$, $CO_2$, $CH_3Cl$, and CO, are indicated (see **Table 3**). Line descriptions are consistent with those in **Figure 3**. Solid gray lines represent SH planets receiving an incident stellar flux of 0.8 solar constants ($S_0 = 1366$ W/m²), positioned between the inner edge and center of their respective host star's habitable zone (HZ), while dashed lines represent SH planets at 0.6 $S_0$, aligned with the center of the HZ.

**Table 3:** Key absorption wavelengths of major atmospheric constituents in Earth-like atmospheres in the *POSEIDON* wavelength range of 0.3–14 micrometers.

| Species | Wavelength (µm) | Reference |
|---------|-----------------|-----------|
| $O_3$ | 0.31–0.34, 0.4–0.85, 4.8, 9.6 | (Serdyuchenko *et al.* 2014) |
| $CO_2$ | 1.4, 1.6, 2.0, 2.7, 4.3, 4.8, 9.5, 10.5 | (Wei *et al.* 2018) |
| $CH_4$ | 0.9, 1.1, 1.4, 1.7, 2.2, 2.3, 3.3, 6.5, 7.1, 7.7 | (Brunetti and Prodi 2015) |
| $H_2O$ | 0.82, 0.94, 1.1, 1.4, 1.9, 2.7, 5.5–7.0 | (Wei *et al.* 2019) |
| $CH_3Cl$ | 3.3, 7.0, 9.7, 13.7 | (Schwieterman *et al.* 2018) |
| $N_2O$ | 2.8, 3.5, 3.9, 4.4, 7.5 | (Plumb and Marshall 1961) |
| CO | 1.6, 2.4, 4.7 | (Stepanov *et al.* 2020) |
| $O_2$ | 0.63, 0.69, 0.76, 0.77, 0.86, 1.06, 1.27, 1.58 | (Catling *et al.* 2018) |

### 3.4. *JWST Simulations and Retrievals*

Simulated JWST observations and atmospheric retrievals have been conducted for all scenarios using the *PandExo* tool, which incorporates modeled observational noise around a *POSEIDON* model (**Figure 5** and **6**). In *PandExo*, noise is simulated by combining photon noise, thermal background from the telescope, instrumental noise, and estimated systematic errors (Batalha *et al.* 2017). Here, we present the results for one specific scenario: the SH planet located in the habitable zone of a 4300K host star, receiving an incident stellar flux of 0.8 solar constants at a distance of 30 parsecs (**Figure 7**). This scenario is particularly notable because it exhibits a relatively large transit depth and prominent spectral features in its transmission spectrum with 0.8 $S_0$ (**Figure 5** and **6**, respectively). Additionally, its semi-major axis does not fall within the range of distances where a planet would be tidally locked, unlike the SH planet orbiting a 3900K host star. Consequently, the remaining example results will focus on this scenario. The corresponding figures for the other scenarios are available in the **Supplementary Materials (Figure S1** through **S7)**.

**Figure 7** displays the retrieval results from simulated JWST observations, generated using the *PandExo* tool. Here we target the 1.27 μm $O_2$ feature with NIRSpec G140H, the 2.3 μm $CH_4$ feature with NIRSpec G235H, the 4.3 μm $CO_2$ feature with NIRSpec G395H, and the 7.5 μm $N_2O$ and 9.6 μm $O_3$ features with MIRI LRS. The number of transits used for each instrument is summarized in **Table 4** and reflects the minimum necessary to accurately constrain the spectral features within their 1σ confidence intervals. The yellow diamonds mark the median values of the binned data points obtained through the retrieval process. The purple curve indicates the median retrieved spectrum. The dark and light purple areas illustrate the 1σ and 2σ confidence intervals for the transit depth at each wavelength, respectively. These confidence intervals are determined from 10,000 random samples drawn from the posterior distribution.

The corresponding corner plot depicting the joint posterior distributions for NIRSpec G235H is shown in **Figure 8**. The diagonal elements of the corner plots illustrate the marginal distributions (histograms) for individual parameters: planetary radius (R), planetary temperature (T), as well as the logarithm of the concentrations of $O_2$, $CO_2$, $CH_4$, $H_2O$, $O_3$ and $N_2O$. The off-diagonal elements show joint distributions of parameter pairs, indicating correlations. Green lines mark the true parameter values used for the simulated data, which correspond to the logarithm of vmrs at $10^{-2}$ bar. Blue lines in the histograms represent the median values obtained from the retrieval. All parameters are successfully retrieved within their respective 1σ confidence intervals. The true values and the corresponding retrieved values for this scenario are summarized in **Table 4**.

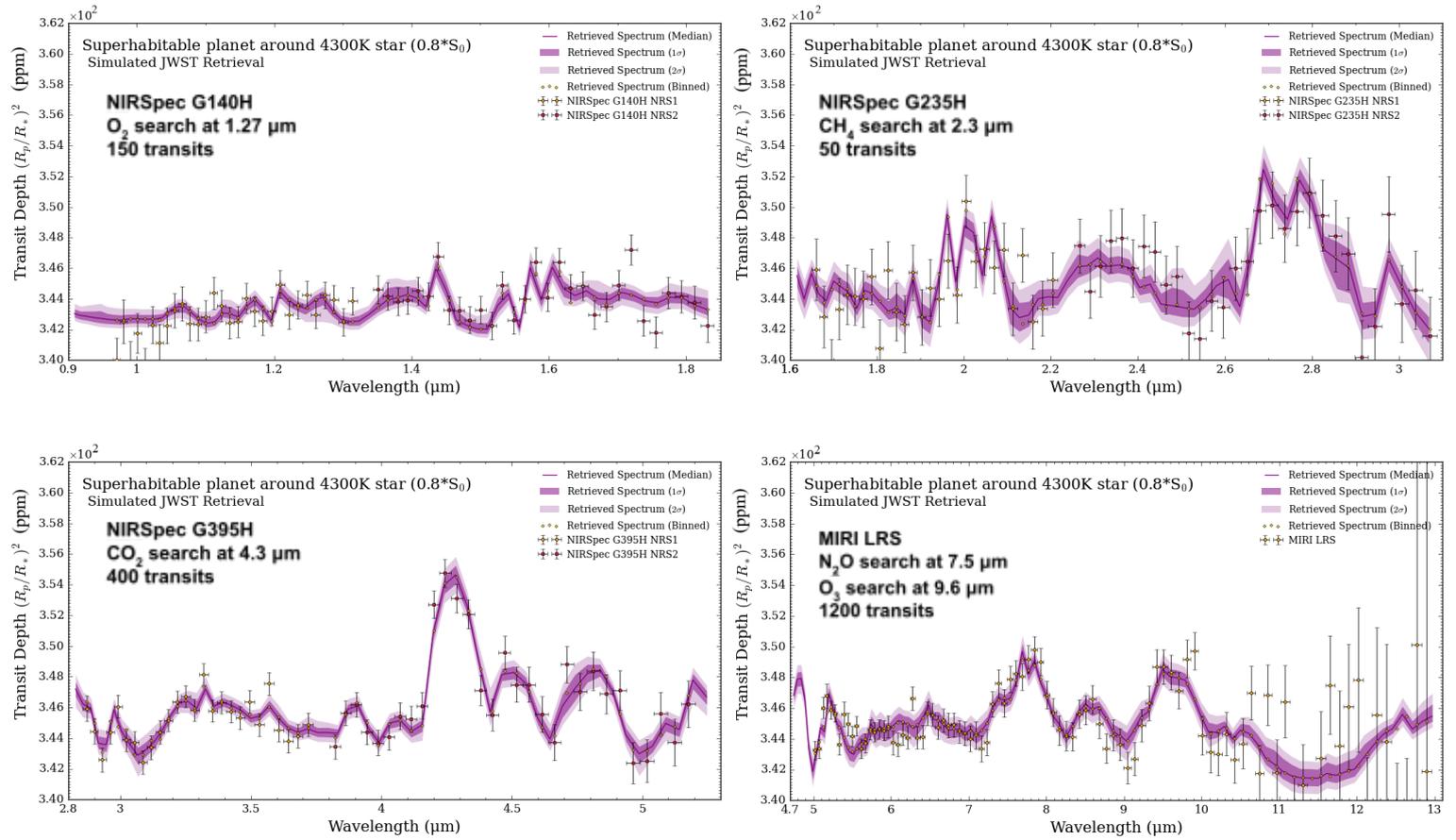

**Figure 7:** Selected retrieval of the simulated James Webb Space Telescope (JWST) observations obtained using the *PandExo* tool. The simulations are for a superhabitable (SH) planet located in the habitable zone (HZ) of a 4300K host star, receiving an incident stellar flux of 0.8 solar constants ($S_0$ = 1366 W/m²) at a distance of 30 parsecs from us. The signal-to-noise ratio (S/N) for observations is proportional to $\sqrt{N_{trans}}$, with $N_{trans}$ denoting the number of transits. Scatter around the *POSEIDON* forward model from the high-resolution NIRSpec instruments G140H (0.97–1.89 μm), G235H (1.66–3.17μm) and G395H (2.87–5.27μm), as well as the low-resolution MIRI LRS instrument (5.0 – 12.0 μm) are shown. The NIRSpec instruments include band gaps at 1.3, 2.2 and 3.8 μm, respectively, which partition the data into the 'NRS1' and 'NRS2' detectors. A high-resolution (R ≈ 10 000) spectrum is generated by the forward model, convolved with the point spread functions (PSF) of the individual JWST instruments and integrated over the corresponding instrument functions to produce binned synthetic model points. The yellow diamonds represent the median values of the binned model points obtained from the retrieval process. The dark and light purple regions represent the 1σ and 2σ confidence intervals, respectively, for the transit depth at each wavelength.

**Table 4:** Retrieval results from simulated James Webb Space Telescope (JWST) observations for superhabitable (SH) planets positioned midway between the center and inner edge of the habitable zone (HZ), receiving an incident stellar flux of 0.8 solar constants ($S_0$ = 1366 W/m²), around stars with temperatures of 3900K, 4300K, and 4900K at a distance of 30 parsecs. Results are also shown for Kepler-62e (1.2 $S_0$) and Earth around the Sun (1.0 $S_0$) at 30 parsecs. Here we target the 1.27 μm $O_2$ feature with NIRSpec G140H, the 2.3 μm $CH_4$ feature with NIRSpec G235H, the 4.3 μm $CO_2$ feature with NIRSpec G395H, and the 7.5 μm $N_2O$ and 9.6 μm $O_3$ features with MIRI LRS. The true values of the log of volume mixing ratios of each chemical species at $10^{-2}$ bar are shown, along with the minimum number of transits needed to constrain these features in each scenario. The corresponding median retrieved values are provided. The observation time (Obs. time) in years, calculated per the planets' orbital periods, is also listed. Parameters with a specific number of transits are retrieved accurately within their 1σ confidence intervals. Parameters denoted by '>' still could not be accurately retrieved within their 1σ confidence intervals, indicating the limitations in detection despite extensive observational data.

| Parameter log(species) | 3900K SH (0.8 $S_0$) | | | | 4300K SH (0.8 $S_0$) | | | | 4900K SH (0.8 $S_0$) | | | |
|---|---|---|---|---|---|---|---|---|---|---|---|---|
| | Number transits | Obs. time [yr] | True value | Retrieved value $^{+1\sigma}_{-1\sigma}$ | Number transits | Obs. time [yr] | True value | Retrieved value $^{+1\sigma}_{-1\sigma}$ | Number transits | Obs. time [yr] | True value | Retrieved value $^{+1\sigma}_{-1\sigma}$ |
| log($O_2$) | 350 | 69 | -0.61 | $-0.54^{+0.10}_{-0.11}$ | 150 | 43 | -0.61 | $-0.55^{+0.15}_{-0.24}$ | 250 | 126 | -0.60 | $-0.67^{+0.15}_{-0.17}$ |
| log($CH_4$) | 100 | 19 | -4.79 | $-4.13^{+0.63}_{-0.78}$ | 50 | 14 | -4.78 | $-4.99^{+0.56}_{-0.44}$ | 100 | 50 | -4.76 | $-4.60^{+0.45}_{-0.48}$ |
| log($CO_2$) | 200 | 39 | -1.54 | $-1.40^{+0.30}_{-0.33}$ | 400 | 115 | -1.54 | $-2.20^{+0.84}_{-0.63}$ | 200 | 101 | -1.50 | $-1.64^{+0.46}_{-0.53}$ |
| log($O_3$) | >1200 | >236 | -6.17 | $-7.12^{+0.31}_{-0.31}$ | 400 | 115 | -6.02 | $-6.70^{+0.77}_{-0.69}$ | 1000 | 506 | -5.83 | $-5.82^{+0.66}_{-0.68}$ |
| log($N_2O$) | 600 | 118 | -5.57 | $-5.51^{+0.56}_{-0.59}$ | 1200 | 345 | -5.56 | $-5.70^{+0.35}_{-0.34}$ | 1500 | 760 | -5.55 | $-5.71^{+0.36}_{-0.35}$ |

| Parameter log(species) | Kepler-62e (1.2 $S_0$) | | | | ME Sun (1.0 $S_0$) | | | |
|---|---|---|---|---|---|---|---|---|
| | Number transits | Obs. time [yr] | True value | Retrieved value $^{+1\sigma}_{-1\sigma}$ | Number transits | Obs. time [yr] | True value | Retrieved value $^{+1\sigma}_{-1\sigma}$ |
| log($O_2$) | >1000 | >334 | -0.60 | $-0.41^{+0.04}_{-0.04}$ | 1000 | 999 | -0.68 | $-0.53^{+0.14}_{-0.10}$ |
| log($CH_4$) | 125 | 42 | -5.08 | $-5.13^{+0.52}_{-0.35}$ | 1700 | 1699 | -5.94 | $-5.69^{+0.35}_{-0.31}$ |
| log($CO_2$) | >3000 | >1002 | -4.29 | $-1.73^{+0.12}_{-0.13}$ | >4000 | >4000 | -3.41 | $-1.24^{+0.11}_{-0.12}$ |
| log($O_3$) | 800 | 267 | -6.19 | $-6.59^{+0.34}_{-0.29}$ | >4000 | >4000 | -5.19 | $-6.86^{+0.25}_{-0.22}$ |
| log($N_2O$) | 800 | 267 | -5.66 | $-5.64^{+0.49}_{-0.36}$ | >4000 | >4000 | -6.84 | $-7.41^{+0.32}_{-0.34}$ |

For orbital period details, used for calculating the total observation time in years, refer to **Table 1(a)**. The number of transits needed to constrain spectral features is generally much higher for the low-resolution MIRI LRS instrument compared to the high-resolution NIRSpec instruments. For example, the 3900K (0.8 $S_0$) scenario requires 100 transits with NIRSpec G235H to constrain the 2.3 μm $CH_4$ feature, but more than 1200 transits with MIRI LRS to constrain the 9.6 μm $O_3$ feature. This discrepancy increases further for the 4300K and 4900K (0.8 $S_0$) scenarios, where the $O_3$ feature requires 400 and 1000 transits, respectively. The 4300K SH planet requires 50 transits to constrain the 2.3 μm $CH_4$ feature, leading to the shortest observational time of 14 years. In general, the 0.8 $S_0$ SH scenarios require significantly fewer transits with NIRSpec to constrain spectral features compared to Earth around the Sun. While the retrievals are well-constrained, it is important to note that the high number of transits and extended observational times required to detect certain spectral features present practical challenges for JWST. For many scenarios, particularly those requiring hundreds or thousands of transits, the total observation times would exceed JWST mission timescales. This limitation suggests the necessity of focusing on the most critical target features and optimal instrument configurations, or exploring the capabilities of next-generation, more sensitive missions.

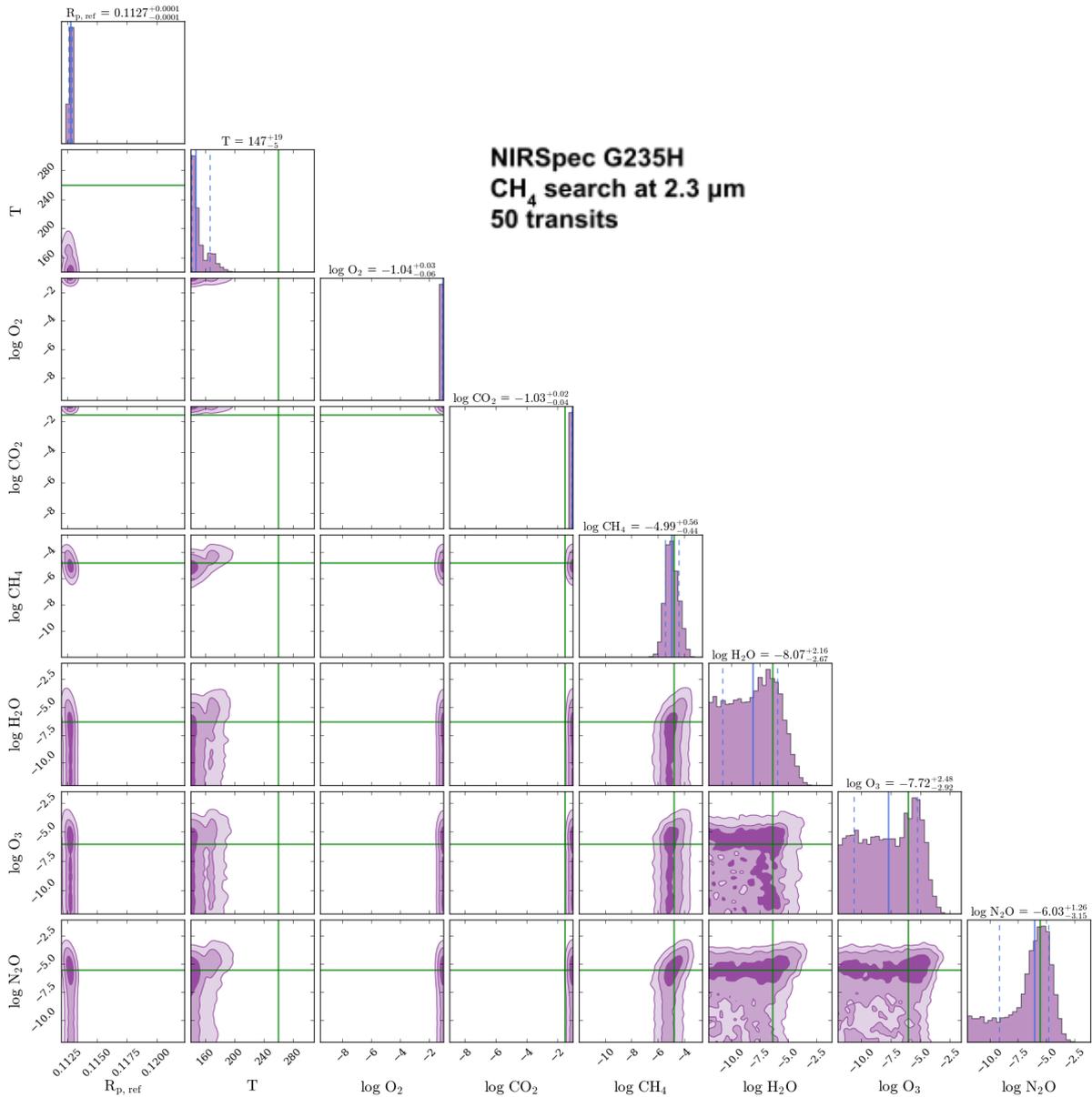

**Figure 8:** Selected posterior distributions from the retrieval analysis of James Webb Space Telescope (JWST) high-resolution NIRSpec G235H instrument data, as shown in **Figure 7**, obtained using the *PandExo* tool with 50 transits. Each diagonal element of the corner plots shows the marginal distribution (histogram) of a single parameter–planetary radius R, planetary temperature T, oxygen $O_2$, carbon dioxide $CO_2$, methane $CH_4$, water vapor $H_2O$, ozone $O_3$ and nitrous oxide $N_2O$–representing the posterior distribution for that parameter. Here the focus is on the 2.3 micrometer $CH_4$ feature, which is constrained with 50 transits to 1σ. The off-diagonal elements show joint distributions of parameter pairs, indicating correlations. Diagonally elongated contours suggest strong correlations, narrow circular distributions imply well-constrained, uncorrelated parameters, elliptical contours indicate well-constrained but correlated parameters, and broad, diffuse distributions point to poorly constrained parameters. The green lines indicate the true values of each parameter used to generate the simulated data (=the log of volume mixing ratios at 10 mbar). The blue lines in the histograms correspond to the median retrieved values.

## 4. Discussion

Previous laboratory experiments have shown that phototrophic organisms can grow under simulated K dwarf radiation; some organisms even demonstrated significantly better growth responses to K dwarf radiation compared to solar radiation (Vilović *et al.* 2024). Research on Earth-based photosynthetic adaptations supports this possibility: for example, chlorophyll *d* and *f*, used by certain organisms, allow photosynthesis under far-red and near-infrared light, which is characteristic of M dwarf radiation (Claudi *et al.* 2020, Battistuzzi *et al.* 2023). Studies on phototrophic extremophiles on Earth show that organisms in low-light or red-light environments often evolve specialized cellular structures and pigments to maximize light absorption (Kiang *et al.* 2007). Since K dwarfs emit more visible light than M dwarfs, the likelihood of achieving similar or even greater productivity under K dwarf radiation is strong. Over geological timescales, organisms on a K dwarf planet could similarly develop pigments optimized for these specific wavelengths, potentially achieving productivity rates comparable to or greater than Earth's phototrophs.

However, there has been a dearth of data simulating superhabitable conditions numerically. **In this study, to our knowledge, we present the first simulated data on superhabitable exoplanetary atmospheres, assess the observability of life on these worlds, identify the most observable types, and estimate optimal observation times.** Caution is needed when interpreting the data, as the study assumes constant mixing ratios and fixed surface conditions. This is due to the limitations of the modeled stellar spectra in the UV region, which may not fully capture all the photochemical processes occurring in the atmospheres of the simulated exoplanets. Incorporating stellar spectra–observed or modeled–that fully capture the UV regime, along with variable boundary conditions, such as biological surface fluxes, could enhance the accuracy of the model results.

### 4.1. *Impact of Stellar Properties on Abundance Profiles and Transmission Spectra*

The general trends of the T–P profiles we obtained (**Figure 3**) align with previous research on simulated K dwarf atmospheres (Grenfell *et al.* 2007, Kaltenegger and Lin 2021). The $H_2O$ profile (**Figure 4**) aligns with the T–P profile, showing that as incident stellar flux decreases from 1.2 $S_0$ for Kepler-62e to 0.6 $S_0$ for the SH planets, water vapor concentrations also decrease. Lower stellar fluxes result in cooler surface temperatures, which reduce the amount of water vapor in the atmosphere. This reduction in water vapor further lowers the temperatures in the T–P profile. In assessing the sensitivity of the *Atmos* model, the largest changes in T–P profiles occur with variations in the incident stellar flux, in combination with higher surface pressures, which significantly impact surface temperature and atmospheric composition. The model is also somewhat sensitive to changes in surface stellar gravity, which can further affect atmospheric pressure and scale height. However, changes in planetary radius and minor changes in other biological gases (e.g., $O_2$, $N_2O$, $CH_4$) have minimal effects on the T–P profiles, relative to the other parameter variations in this study, indicating they play a lesser role in influencing atmospheric structure.

### 4.2. *Assessing UV Sensitivity with Dynamic Boundary Conditions*

To evaluate the combined effects of reduced metallicity and a realistic UV signal on planetary transmission spectra, we calculated the percent differences between two scenarios: a modern Earth atmosphere around a 4300K star receiving 1 solar constant using the modeled PHOENIX spectrum—with only the very small UV flux emerging from the black body spectrum—and a metallicity of Z=0; and a modern Earth atmosphere in orbit around HD 85512 (~4300K), also receiving 1 solar constant, using an observed MUSCLES spectrum available in the *Atmos* code. The observed HD 85512 spectrum captures the full FUV and NUV fluxes and has a metallicity of $Z=−0.5$. For accurate comparison, both scenarios were standardized with a stellar radius of 0.62 $R_{\odot}$ and a planetary radius of 1 $R_{\oplus}$.

In this study, we simulated a modern Earth-like atmosphere around these stars, using surface flux boundary conditions instead of fixed mixing ratios to accurately capture UV sensitivity. This approach allows surface fluxes to adjust dynamically in response to UV-induced chemical changes, unlike fixed mixing ratios, which assume constant concentrations and cannot fully capture atmospheric sensitivities (Segura *et al.* 2003, 2005, 2007, Domagal-Goldman *et al.* 2011, 2014, Arney *et al.* 2017, Arney 2019). For each scenario, we calculated transmission spectra using the *POSEIDON* model and represented the results as absolute transit depth (in ppm) and differential transit depth (change from a baseline value, highlighting smaller-scale variations). We then calculated percent differences between the two transmission spectra using: Percent Difference = [ $|A−B|$ / (A+B)/2 ] × 100, where A and B are transit depth values for each wavelength in the two scenarios. As shown in **Figure S8** of the **Supplementary Materials**, we found distinct percent differences between the absolute and differential transit depths. For absolute transit depths, the percent difference is 0.21% ± 0.14%, reflecting the large baseline values (hundreds of ppm) that limit sensitivity to minor differences. By contrast, differential transit depths show a larger percent difference of 11.45% ± 11.03%, as their smaller baseline values make them more sensitive to small variations. Thus, for the same absolute difference, the percent difference is larger when differential transit depths are smaller. In conclusion, the impact of UV on transmission spectra is small to moderate, depending on whether percent differences are calculated for absolute or differential transit depths.

### 4.3. The Role of $CO_2$ and the Spectral Energy Distribution of K Dwarfs

In general, the SH scenarios in **Figure 3** do not exhibit a stark temperature inversion around $10^{-1}$ bar, like the modern Earth around the Sun and Kepler-62e. This is primarily because the SH cases have much more $CO_2$ (80 and 1500 PAL compared to 1 and 0.1 PAL, respectively) in the atmosphere, which weakens the relative importance of $O_3$ in the radiative budget of the atmosphere. At these high levels, $CO_2$ plays a crucial role in shaping the T–P profile of the atmosphere. Thus, our simulations highlight the importance of $CO_2$ in the radiative budget of superhabitable planets. Additionally, the spectral energy distribution of K dwarf stars differs from that of the Sun. Compared to the Sun, K dwarf stars emit more long-wave radiation and less short-wave radiation. As a result, less short-wave radiation is available to be absorbed by $O_3$ in the upper atmosphere, which means there is less potential for creating a temperature inversion. Kepler-62e, compared to the other three K dwarfs in this study, receives the highest stellar incident flux of 1.2 $S_0$, leading to an increased absorption by $O_3$ in the upper atmosphere and a moderate temperature inversion between $10^{-2}$ and $10^{-3}$ bar. Overall, the T–P profiles of the SH planets align with the findings of Manabe and Wetherald, who conclude that higher concentrations of $CO_2$ lead to warmer surface temperatures and cooler stratospheric temperatures (Manabe and Wetherald 1967). Moreover, Kepler-62 has a lower metallicity ($Z=-0.5$) compared to the other stars in this work. Metal-rich stars emit less UV radiation than metal-poor stars, leading to lower $O_3$ production and, paradoxically, less UV protection (Shapiro *et al.* 2023). However, our results show Kepler-62e, orbiting a metal-poor star relative to the Sun, exhibits less $O_3$ in the upper atmosphere. This discrepancy might be due to Kepler-62e receiving more stellar flux than the other scenarios.

### 4.4. Habitable Zone Location and its Implications for Observability & Habitability

The choice of a SH atmosphere with a surface pressure of 2 bar is justified, as research shows that life forms on Earth can adapt to high-pressure environments, suggesting the potential for extraterrestrial life in similar or more moderate conditions (Ono *et al.* 2016). While extremely high pressures (significantly above 2 bar) can reduce respiration rates in certain plants, a combination of moderate pressures (up to 2 bar) and elevated $CO_2$ levels can enhance net photosynthesis (Wong 1979, Takeishi *et al.* 2013).

With respect to the localization in the stellar habitable zone, previous hypotheses on SH planets propose that a planet in orbit at the center of the HZ of its host star could be considered superhabitable (Heller and Armstrong 2014). This is because these planets are less likely to transition into a moist or runaway greenhouse state as the host-star's luminosity changes throughout its main-sequence lifetime. However, here we find that the planets located at the midway point between the inner edge and the center of the HZ (0.8 $S_0$) are preferable both in terms of observability and maintaining a temperate surface with more moderate $CO_2$ concentrations compared to the 0.6 $S_0$ scenarios (HZ center) (**Figure 2** and **6**).

For example, our results show that maintaining a surface temperature of 294K at 0.8 $S_0$ requires 80 PAL of $CO_2$ (~3% by volume). However, when the incident stellar flux is reduced further to 0.6 $S_0$, we observe a significant temperature drop, requiring approximately 1.3 orders of magnitude more $CO_2$ (about 1500 PAL, or 60% by volume) to maintain the same surface temperature. This is consistent with the Faint Young Sun Paradox, which refers to the apparent contradiction between geological evidence for a warm, early Earth and the astrophysical expectation that the Sun's luminosity was 25-30% lower than present (Charnay *et al.* 2013, Wolf *et al.* 2018, Heller *et al.* 2021). The warm conditions on early Earth are generally attributed to high levels of atmospheric $CO_2$ resulting from intense early volcanism. Extreme concentrations of $CO_2$ (i.e., >50 PAL), tested on Earth organisms on non-geological timescales, however, have been shown to have negative effects on plant productivity in some species (Grotenhuis *et al.* 1997). Extremely high $CO_2$ levels can also impact other aspects of the environment, such as soil chemistry, which in turn affect (plant) life. On the other hand, the Carboniferous Era, known for the highest biomass and biodiversity on Earth to date, was marked by up to 10 PAL of $CO_2$ and 28% $O_2$ in the atmosphere (Vilović *et al.* 2023). This suggests that if there is a sufficient amount of $O_2$, but below the spontaneous combustion limit of approximately 30%, moderately high $CO_2$ concentrations can be tolerated and may even be preferred by plant life. Consequently, the 80 PAL $CO_2$ necessary for the 0.8 $S_0$ scenarios might be moderate enough to support a thriving biosphere, without having toxic effects. This is particularly true when considering that life forms on such a planet could adapt and evolve over geological timescales to tolerate these $CO_2$ concentrations.

### 4.5.  *Observability of Superhabitable Planets with Current Telescopes*

The transit depths of the solar and K dwarf results in this study match those in literature (Kaltenegger and Lin 2021). In terms of observability, although a 0.8 $S_0$ SH planet orbiting a 4300K star has lower overall $O_3$ contents compared to Earth, its 9.6 μm $O_3$ feature is more pronounced than that of modern Earth. Furthermore, the 0.8 $S_0$ scenarios exhibit stronger spectral features compared to the 0.6 $S_0$ scenarios. This is because a decrease in planetary temperature reduces the scale height (H) of the planet's atmosphere, assuming the mean molecular mass of atmospheric gases (m) and the acceleration due to gravity (g) remain constant: $H = kT / mg$, where k is the Boltzmann constant, and T is the temperature of the atmosphere. The 0.6 $S_0$ scenarios have the coolest T–P profiles, despite having the same surface temperatures of ~294K (**Figure 3**). This results in a smaller scale height and therefore less prominent spectral features. The SH planet around the 3900K host star (~0.5 $M_\odot$) most likely falls within the tidal locking radius (see Figure 16 in (Kasting *et al.* 1993)). Our results suggest that a planet like Kepler-62e might receive excessive stellar flux, making its surface too hot. Moreover, it is unclear whether its size of more than 1.6 Earth radii might inhibit plate tectonics which are crucial for the recycling of nutrients. **Thus, our results suggest** that a SH planet with 0.8 $S_0$ around a 4300K star is the optimal choice when considering both observability and surface habitability (e.g., surface temperatures, $CO_2$ levels, etc.).

When considering JWST instrument noises at 30 parsecs, the observational prospects for K dwarfs show certain advantages relative to other targets. Simulated observations indicate that detectable biosignature pairs, such as $O_2$ and $CH_4$ (1.27 μm and 2.3 μm, respectively), can be observed in the NIR with 150 and 50 transits for the 0.8 $S_0$ SH planet around a 4300K star, respectively (**Table 4**). This corresponds to an observation time of 43 years, respectively. In contrast, constraining the 9.6 μm $O_3$ feature would require 400 transits with the low-resolution MIRI LRS instrument. Hence, the strong biosignature pair $O_3$+$CH_4$ in the NIR to IR range (3–14 μm) would require a significant 115 years of observation time. Compared to this study, research focusing on smaller M dwarfs like AD Leo (~3500K and ~0.4 $R_\odot$) at a distance of 30 parsecs shows that approximately 40 transits are needed to constrain the 2.3 μm $CH_4$ feature, whereas more than 1000 transits are necessary to constrain the 1.27 μm $O_2$ feature (Wunderlich *et al.* 2019, Gebauer *et al.* 2021). While Kepler-62e is considered a strong candidate for future biosignature searches because of its favorable planetary characteristics, this study reveals that detecting $O_3$ and $O_2$ features with the JWST would require a significant 800 and >1000 transits, respectively. It is therefore important to recognize that though K dwarfs are generally less active than M dwarfs, stellar variability over these extended observation periods presents challenges. Co-adding decades' worth of data may still capture fluctuations due to stellar activity, affecting the stability and reliability of the observed spectral signal and complicating detection of weak features. Furthermore, for many scenarios, especially those requiring hundreds to thousands of transits, the total observation time would exceed typical JWST mission timescales. This limitation emphasizes the need for prioritizing target features and instrument settings or considering future, more sensitive missions to feasibly explore such planetary scenarios.

### 4.6. *Relevance to Future Missions and Observational Strategies*

A telescope with a mirror twice the size of JWST–such as NASA's Habitable Worlds Observatory (HWO)–would need only one-quarter of the transits to constrain spectral features (Burns 2024). This scaling relationship between mirror diameter and S/N arises because larger mirrors collect more photons, and S/N grows linearly with mirror diameter while increasing with the square root of the number of transits. Consequently, to achieve the same S/N, doubling the mirror size reduces the required transits by a factor of four. The HWO flagship mission is planned to launch after the Nancy Grace Roman Space Telescope (scheduled for mid-2027), with a specific focus on the atmospheric characterization of potentially habitable exoplanets, particularly those around nearby stars. While HWO's larger mirror provides an improvement over JWST, these observations may still face practical challenges, including long observation times for individual objects and potential oversubscription.

In this progression of next-generation telescopes, the Extremely Large Telescope (ELT), with its unprecedented 39-meter diameter mirror, represents a major step toward observational feasibility for single-object studies. ELT's ANDES spectrograph focuses on detecting the presence of biomarkers such as $O_2$ and $CH_4$ within the 0.35–2.40 μm wavelength range (Marconi *et al.* 2024). Based on our calculations of the scaling between mirror diameter and the required number of transits, the ELT's large mirror could constrain the $CH_4$ and $O_2$ features in as few as 1 to 4 transits, or 0.3 to 1.2 years of observation time, respectively.

# 5.   Conclusion

Our study identifies the optimal distance for superhabitability around K dwarf stars. We find that planets positioned at the midpoint between the inner edge and center of the habitable zone, where they receive 80% of Earth's solar flux, are more conducive to life. This contrasts with previous suggestions that planets at the center of the habitable zone—where our study shows they receive about 60% of Earth's solar flux—are the most favorable for life (Heller and Armstrong 2014). Planets at the midpoint between the center and the inner edge need less $CO_2$ for temperate climates and are more observable due to their warmer atmospheric temperatures and larger atmospheric scale heights. We conclude that a superhabitable planet orbiting a 4300K star with 80% of the solar flux offers the best balance of observability and habitability.

Unlike tidally locked planets around 3900K stars, these planets have larger transit depths without excessive surface temperatures as are potentially present on Kepler-62e. For biosignature detection with the James Webb Space Telescope (JWST) at 30 parsecs, such a planet requires 150 transits (43 years of observation time) to constrain the $O_2$–$CH_4$ biosignature pair, compared to 1700 transits (1699 years) for Earth around the Sun. Moreover, only 50 transits are needed to observe the prominent $CH_4$ feature at 2.3 microns.

Observations of this nature, however, are not feasible within JWST's planned mission duration and may remain impractical even with NASA's Habitable Worlds Observatory (HWO). While the JWST has opened new possibilities for exoplanetary biosignature detection, our findings underscore the necessity of next-generation telescopes for a comprehensive study of superhabitable exoplanets. Future missions with larger mirror sizes, such as the Extremely Large Telescope (ELT), would significantly reduce the required number of transits and associated observation times. This study, alongside projects like the KOBE experiment (Lillo-Box *et al.* 2022), highlights key exoplanetary targets for such future atmospheric characterization missions.

**Acknowledgements**

The first author would like to thank the scholarship organization *Studienstiftung des deutschen Volkes* without which this work would not have been possible. The authors also thank Swaroop Ramakant Avarsekar and Dibya Bharati Pradhan from NISER, as well as Sandra Bastelberger and Ryan MacDonald for extensive and fruitful discussions concerning the functionalities of *Atmos* and *POSEIDON*. JG acknowledges support from the SERB SRG Grant SRG/2022/000727-G for this work. RH acknowledges support from the German Aerospace Agency (Deutsches Zentrum für Luft- und Raumfahrt) under PLATO Data Center grant 50OO1501.

**Funding**

*Studienstiftung des deutschen Volkes* doctoral scholarship (IV)
SERB SRG Grant SRG/2022/000727-G (JG)
PLATO Data Center grant 50OO1501 (RH)

**Author contributions**

I.V. was involved in the conceptualization, methodology, investigation, visualization, formal analysis, funding acquisition and writing of the original draft as well as the review and editing. R.H and J.G. were involved in the formal analysis, supervision and editing of the manuscript. M.V.S. was involved in the selection of an existing planet model for simulations.

**Data and materials availability**

All data generated or analyzed during this study are included in this published article and their sources are cited where applicable.

**Competing interests**
The authors declare no financial or commercial conflict of interest.
**Correspondence and requests for materials should be addressed to**
Iva Vilović
**Supplementary Materials**
Yes

## Supporting Information

Additional supporting information is available in the online version of this article at the publisher's website or from the author.

Supplementary Text and Equations (1) to (13)

Table S1

Figures S1 to S8